# Modular interface for efficient optical readout of diamond quantum memory at cryogenic temperatures via single-mode optical fibers

Akira Kamimaki[1,2,*], Yuhei Sekiguchi[1,2], Daisuke Ito[3], Taichi Fujiwara[3], Toshiharu Makino[1,2,4], Hiromitsu Kato[1,2,4], Hideo Kosaka[1,2,3]

[1]Institute of Advanced Sciences (IAS), Yokohama National University, 79-5 Tokiwadai, Hodogaya, Yokohama, 240-8501, Japan
[2]Quantum Information Research Center (QIC), Institute of Advanced Sciences, Yokohama National University, 79-5 Tokiwadai, Hodogaya, Yokohama, 240-8501, Japan
[3]Department of Physics, Graduate School of Engineering Science, Yokohama National University, 79-5 Tokiwadai, Hodogaya, Yokohama, 240-8501, Japan
[4]Advanced Power Electronics Research Center, National Institute of Advanced Industrial Science and Technology (AIST), 1-1-1 Umezono, Tsukuba, Ibaraki, 305-8568, Japan
*kamimaki-akira-sx@ynu.ac.jp

**Abstract**

**Efficient quantum devices across various physical systems have been rapidly developed for entanglement-based quantum repeaters and spin–photon conversion; however, far less attention has been paid to standardizing platforms through quantum memory optical interfaces. We present a modular interface for color centers in diamond that is structurally isolated from device-and temperature-related variation. Despite a more than 100-fold reduction in confocal volume, we achieve highly efficient photon collection through single-mode optical fibers, including zero-phonon line spectroscopy, at both room and cryogenic temperatures. These results establish a standardized minimal NV-center-based platform and pave the way for construction of scalable quantum infrastructure.**

The construction of scalable quantum networks [1-3] has become a core challenge in modern quantum technology, involving not only quantum key distribution (QKD) but also entanglement-based long distant communication [4,5] and distributed quantum computing [6,7] are being actively developed. The advanced applications are built upon quantum interconnection layers, including quantum repeaters [8-14], spin–photon interfaces [15,16], and media converters [7,17-19], which together enable the coherent generation, storage, and transfer of quantum states between spatially separated or heterogeneous quantum nodes. In each physical system, entanglement swapping based on Bell-state measurements has been successfully used to connect remote nodes [8,9,20], while in heterogeneous systems, interfaces based on various operating principles have been reported [7].

A physical system underpinning these layers is expected to serve as a scalable platform, which requires not only high quantum performance but also efficient optical interconnection with an internal memory. One of the most promising candidate systems is the nitrogen-vacancy (NV) centers in diamond, which offer optically controllable spins, reliable quantum memory retention, and excellent integration with photonic networks [22-26]. While experimental demonstrations of Bell-state measurements and entanglement swapping between NV nodes have been driving the development of photon-interference-based quantum repeaters [10-15,20], new advances have also been reported, such as absorption-and emission-based technologies for spin–photon interface elements [16,19,26-29], and electric field control of orbital states in neutral-charge $NV^0$ centers for optical–microwave interfacing [30,31].

On the other hand, the NV-center based platform still faces optical challenges. Similar to the case in many solid-state hosts, prominent total internal reflection degrades photon extraction efficiency [32], and the small Debye–Waller (DW) factor (~4%) limits the zero-phonon line (ZPL) emission yield [33]. There is a growing focus on cavity-enhanced photonic devices and fiber-coupled module implementations for NV centers [24,25,34] as with various quantum systems [35-37], which are prominent strategies for high-speed quantum communication and efficient quantum conversion. In addition to the high efficiency, however, the practical implementation of these devices requires wide bandwidth, robustness in varied environment such as thermal cycling and mechanical vibrations, and integration of heterogeneous materials [38, 39]. These requirements are fundamental challenges across different physical systems and represent the direction for standardization and scalable implementation.

Departing from conventional device-centered approaches, this work proposes a concept for modularizing the quantum memory optical interface in an NV-center-based platform. Specifically, we design a modular interface for confocal microscopy between the objective lens and single-mode optical fibers (SMFs) and demonstrate efficient photon collection from the NV center regardless of operating temperature. Precise alignment of minimal optical components within the isolated housing not only reduces the volume to less than 1/100 but also allows for simultaneous highly efficient optical connection and mechanical isolation of the devices, even in a cryostat (Fig. 1). The complementary design is different from the conventional modules [37] and instead focuses on providing an optical readout function independent of the device level and quality. As in previous demonstrations of the NV-center-based remote entanglement generation [10-12], we utilized hemispherical solid immersion lenses (SILs). The device and the objective lens in front of the modular interface are arranged in a conventional horizontal configuration (Fig. 1(d)). The integrated modular system, combining a variety of devices and the modular interface, results in a platform that is adaptable to various physical systems.

Since the NV centers are utilized at both room temperature (RT) and cryogenic temperatures, the modular interface is designed as a minimal setup optimized for each environment. There are two combinable modules: a 2-port lens-filter module for non-resonant photoluminescence (PL) measurements with green laser excitation, and a 5-port filter module for resonant photoluminescence excitation (PLE) measurements at cryogenic temperatures and multi-node connection. The 2-port lens-filter module allows either direct coupling with an objective lens (for RT use; Fig. 1(c)) or indirect coupling inside the cryostat via fixtures and flanges (Fig. 1(d)). Note that even for cryogenic experiments, the modular system remains outside the cryostat. All optical paths are SMF coupled, and the loss at the designated wavelength remains below 0.5 dB across all connection paths. In addition, the collimators at the edge of all ports are attached and secured with adhesives, which leads to long-term stability of the optical axis.

To compare the performance of the modular system with that of a conventional free-space optical system, we measure non-resonant PL excited by a green (515 nm, continuous-wave) laser, detected by an avalanche photodiode (APD). Figure 3(a) shows the excitation power dependence of PL counts for both RT and cryogenic (5 K) conditions, comparing the modular and free-space systems. The PL count $C(P)$ with various excitation power $P$ follows the saturation curve: $C(P) = C_0 \cdot P/(P + P_0)$, where $C_0$ and $P_0$ are the saturation count and power, respectively [32]. The background due to setup is subtracted for all data. All PL curves in Fig. 3(a) are thus normalized by the saturation count obtained in the free-space system at RT. In addition, the saturated counts for the ZPL and phonon sideband (PSB) at 5 K are also plotted, which are spectrally separated by coupling with the 5-port filter module. Note that the PL count at 5 K typically decreases to about 70% of RT even in the free-space system. Meanwhile, the fiber-connection loss between these modules is excluded, and the obtained ZPL counts relative to the PSB counts is about 4.9%. The count ratio being comparable to the known DW factor (~4%) [33] indicates that the modular system achieves not only efficient photon collection with the 2-port module but also high spectral efficiency for ZPL separation with the 5-port module. This is particularly important because the ZPL corresponds to the direct transition of the NV center—that is, the optical readout from the quantum memory—and plays a critical role in the efficient remote entanglement generation. The NV center is positioned at the center of the SIL to enhance interaction with photons (Fig. 3(b)) and the corresponding PL map using the modular system is clearly observed (Fig. 3(c)). Here, the PL map is acquired by sweeping the piezo actuator integrated into the objective lens in combination with the 2-port module. Figure 3(d) shows the saturated PL counts of various SILs (2 or 10 μm diameter) obtained by the free-space and (2-port lens-filter) modular systems; the SMF coupling efficiency in ZPL spectroscopy is also comparable, as shown above. In all devices, the modular system achieves SMF coupling efficiencies of either about 50% (solid line) or about 100% (dashed line) compared with the free-space system, which uses multi-mode detection. In general, the SMF coupling efficiency in conventional free-space systems is often limited to around 50% [10], indicating that the modular system serves as a highly efficient optical readout interface for quantum memories.

As a basic demonstration of coherent spin manipulation, resonant PLE measurements using the modular system are also performed at 5 K. The NV center is excited by a resonant (637 nm, continuous-wave) laser, and the PSB photons are spectroscopically detected through the 5-port filter module. A typical PLE spectrum across a broad frequency sweep of up to 20 GHz is exhibited in Fig. 4(a), showing a significant signal-to-noise ratio. The $E_y$-resonant excitation power dependence is also measured, and the saturation count obtained by fitting (using the same curve as for non-resonant excitation) is about 720 kcps, as shown in Fig. 4(b). Although the saturation count is lower than that in the previous report (1100 kcps) [23], this is mainly due to the difference in charge initialization fidelity and device quality. As shown in the inset of Fig. 4(b), initialization is performed using only a non-resonant green laser (~70% fidelity) and the device (SIL-D) is selected considering the effect of spectral diffusion, which becomes critical under resonant excitation, caused by surface processing damage due to the small SIL diameter. Therefore, the collection efficiency obtained here is comparable to, or potentially exceeds, that of the previous report. In addition, the excitation laser is pulsed (50 ns) to prevent relaxation into dark spin states ($|m_s\rangle = |\pm 1\rangle$), and the polarization is optimized using a fiber polarization controller (FPC) in front of the module. Note that although the demonstration requires multiple FPCs—one for each excitation laser—in addition to the two modules, the total system volume remains less than 1/100 that of the free-space system.

These findings highlight the potential of the modular system to achieve efficient and robust photon collection under practical conditions. To achieve this, the module is designed by treating the NV center as an ideal point-source emitter, ensuring that the beam passing through the objective lens, corresponding to a finite solid angle from the emitter, is optimally coupled into the SMF. As noted earlier, the SMF coupling efficiency is typically limited to about 50% in conventional free-space systems, mainly because standard optical components compound the wavefront distortions of the beam emitted from the emitter. In fact, in previous remote Bell experiments using NV centers in SILs, deformable mirrors and 4f optics have been incorporated as the compensating system, and the wavefront reconstruction improved the SMF coupling efficiency of the ZPL emission to nearly twice its original value [10]. The results shown in Fig. 3(d), where the efficiencies are observed at two distinct levels, about 50% (solid line) and 100% (dashed line), indicate that the primary origin of wavefront distortion in our system lies not in the optical components within the modules, but rather in the off-centered position of the NV center in the SIL.

For remote entanglement generation using multiple NV centers, it is important to achieve not only high collection efficiency but also high resonant excitation efficiency. From the fitting in Fig. 4(b), we obtain $P_{\text{sat}} \sim 100\,\mu\text{W}$. In the same way as for collection efficiency, device optimization could reduce the $P_{\text{sat}}$ by about half (~ 50 μW). Since the previous excitation efficiency [23] was 10–20 μW, the maximum excitation efficiency estimated for the module does not yet reach it. Although this study is currently focusing on optimization of the collection efficiency, optimization of the barrel length (including fixtures and flanges) in the module can in principle deliver high excitation efficiency as well, demonstrating that the design concept remains robust for broader applications.

Finally, we summarize how the developed modular interface can contribute to remote entanglement generation using NV centers. By employing NV centers as quantum memories, we demonstrated highly efficient detection of ZPL photons through SMFs under both resonant and non-resonant excitation, while reducing the confocal setup between the objective lens and SMFs to less than 1/100 of its original volume. This compact configuration serves as an optical package robust against device-specific properties and thermal-cycle-induced misalignment, enabling highly efficient and stable telecom-convertible optical readout from diamond quantum memories. Furthermore, the results also address issues raised in the context of constructing quantum interconnection layers and infrastructures [2,3], namely the development of modular optical interfaces beyond device-level engineering and operability across heterogeneous platforms. In this sense, the designed and demonstrated concept provides a practical advancement toward quantum infrastructure, thereby highlighting the often-overlooked role of platform development in tandem with the rapid growth of the current quantum technologies.

**Acknowledgments**

H. Kosaka acknowledges the funding support from Japan Science and Technology Agency (JST) Moonshot R&Dgrant (JPMJMS2062) and JST CREST grant (JPMJCR1773). H. Kosaka also acknowledges the Ministry of Internal Affairs and Communications (MIC) for funding, R&D for construction of global quantum cryptography network (JPMI00316), R&D of ICT Priority Technology Project, and the Japan Society for the Promotion of Science (JSPS) Grants-in-Aid for Scientific Research (20H05661, 20K20441, 25H0083050).

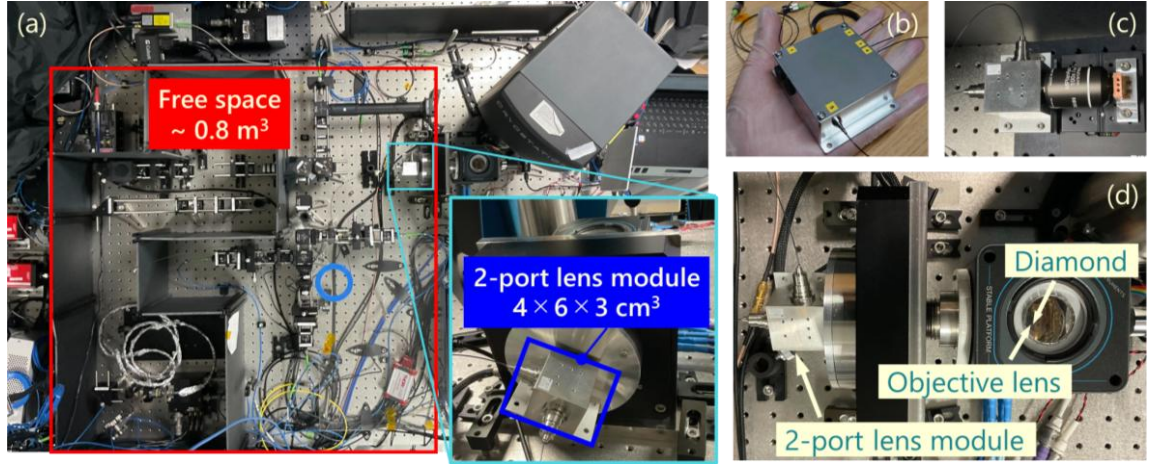

Fig. 1 Compact optical modular interface for nitrogen vacancy (NV) centers in diamond. (a) Size comparison between a conventional free-space confocal microscope and the main part of a single-mode optical-fiber (SMF) coupled confocal microscope (2-port lens-filter) module, achieving a total volume ratio of less than 1/100. (b) Additional (5-port filter) module for resonance excitation and zero-phonon line (ZPL) spectroscopy, which is the same palm size ($6 \times 6 \times 3$ cm$^3$) as the 2-port lens-filter module. (c) Direct

connection between the 2-port lens-filter module and objective lens. The diamond is fixed to a copper base. (d) Integrated connection between 2-port lens-filter module and cryostat-inserted objective lens, which is optically connected without the need for alignment.

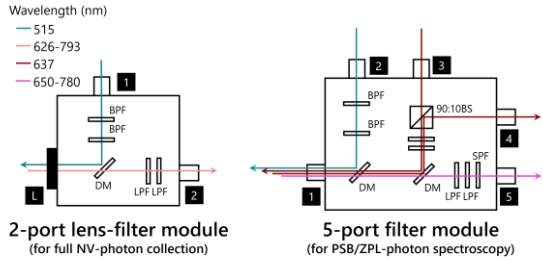

Fig. 2 Internal configuration of the modules designed for highly efficient focusing, spectroscopy and the SMF coupling in the band of the NV center with chromatic aberration suppressed. (Left) The 2-port lens-filter module, which has paths for non-resonant excitation (port 1), the full-wavelength photoluminescence (PL) detection (port 2), and direct or indirect connection to an objective lens (L; Olympus MPLAPON100X, NA 0.95). (Right) The 5-port filter module, which has passes for non-resonant and resonant excitation (ports 2 and 3), spectroscopy of the PL into ZPL (port 4) and phonon sidebands (PSB, port 5), and the device side (port 1). The beam splitter is polarization-independent, and all the SMFs selected for PL detection are pure silica fibers (Thorlabs, S630HP) to minimize fluorescence.

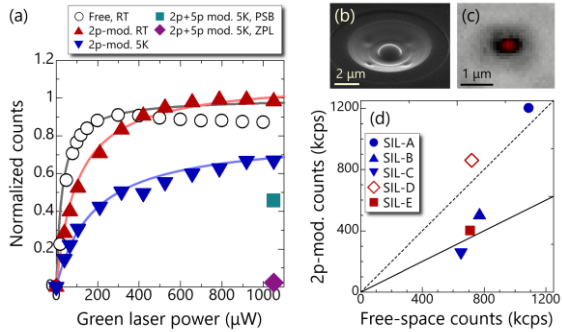

Fig. 3 PL measurements of the NV center in a hemispherical solid-immersion lens (SIL) device with non-resonant excitation using the modular system. (a) Excitation power dependence using the 2-port lens-filter module (2p-mod.). For comparison, data obtained with a free-space system (Free, RT) are also included, all normalized to the saturated count. In addition, the PSB and ZPL counts obtained by coupling a 5-port filter module (2p+5p mod.) at 5K are also included. (b) Scanning electron microscope (SEM) image of a typical hemispherical SIL (2-μm diameter). (c) The PL map obtained by 2p-mod. (d) Saturated counts obtained using the 2p-mod compared with free-space system for various SILs at RT. SIL-A to-C (D and E) are 2(10) μm in diameter. Solid (dashed) lines indicate 50% (100%) counts to the free space. Note that all measurements in (a) are performed by SIL-D.

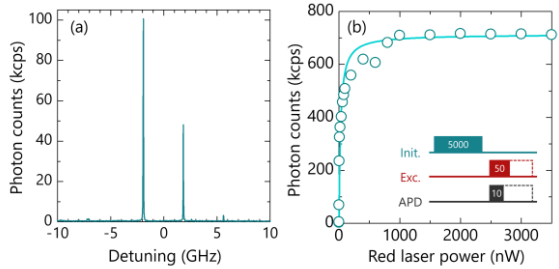

Fig. 4 Photoluminescence excitation (PLE) measurements of the NV center at 5 K using the modular system, detected as SMF-coupled PSB counts by the 5-port filter module connection. (a) Typical PLE spectrum. (b) Saturation curve at $E_y$-resonance frequency where the symbols are measurements and the solid line is the fitting. In (a), the pulse widths of the excitation laser (Exc.) and gating (APD) are both set to 1000 ns, while in (b) they are 50 and 10 ns, respectively.